\documentclass[aps,
amsmath, amssymb,
reprint,
superscriptaddress
]{revtex4-1}

\usepackage{graphicx}
\usepackage{dcolumn}
\usepackage{bm}

\begin{document}

\title{Stimulated Four-Wave Mixing in Linearly Uncoupled Resonators}

\author{K. Tan}
\affiliation{Xanadu, Toronto, ON, M5G 2C8, Canada} 
\author{M. Menotti}
\affiliation{Xanadu, Toronto, ON, M5G 2C8, Canada} 
\author{Z. Vernon}
\affiliation{Xanadu, Toronto, ON, M5G 2C8, Canada} 
\author{J. E. Sipe}
\affiliation{Department of Physics, University of Toronto, 60 St. George Street, Toronto, Ontario M5S 1A7, Canada}
\author{M. Liscidini}
\email{marco.liscidini@unipv.it}
\affiliation{Dipartimento di Fisica, Universit\`{a} degli studi di Pavia, Via Bassi 6, 27100 Pavia, Italy}
\author{B. Morrison}
\affiliation{Xanadu, Toronto, ON, M5G 2C8, Canada}

\begin{abstract}
We experimentally demonstrate stimulated four-wave mixing in two linearly uncoupled integrated Si$_3$N$_4$ micro-resonators. In our structure the resonance combs of each resonator can be tuned independently, with the energy transfer from one resonator to the other occurring in the presence of a nonlinear interaction. This method allows flexible and efficient on-chip control of the nonlinear interaction, and is readily applicable to other third-order nonlinear phenomena.
\end{abstract}

\maketitle

Nonlinear optics plays a central role in the operation of many integrated photonic components, enabling the generation, amplification, and novel manipulation of light. Indeed, a key advantage of working with integrated devices is the possibility of greatly increasing the efficiency of nonlinear optical processes, thanks to the electromagnetic field enhancement typical of micro-resonators \cite{ferrera2008}.  One might expect that a single micro-ring, one of the simplest resonant structures, would be optimal for enhancing nonlinear interactions by providing large field enhancements over a comb of resonances covering a wide range of frequencies. Yet, in some situations, undesired nonlinear phenomena are enhanced along with those of interest. For example, in using a micro-ring for optical parametric oscillation there can be significant power-dependent resonance shifts arising from self- and cross-phase modulation (SPM, XPM) \cite{MatskoPRA2005,Razzari201041,Levy201037}, leading to a detuning of fields from their resonance. In order to compensate for these effects, difficult waveguide dispersion engineering is usually required. This suggests that slightly increasing the complexity of the structure to improve its flexibility might be a practical and advantageous strategy.

One possible approach takes its inspiration from electronic systems such as molecules or crystals, in which the electronic properties are strongly dependent on the spatial arrangement of the constituent elements. Similarly, in integrated optics one can work with a structure composed of two or more linearly coupled resonators to exploit the super-modes of a ``photonic molecule'' or a photonic crystal. This approach is very powerful and has been particularly effective for controlling nonlinear phenomena, with examples ranging from photonic molecules \cite{azzini2013,gentry2014,popovic2015,Miller201521527}, in which the spectral position of the structure resonances is engineered to satisfy the constraint of energy and momentum conservation required by the nonlinear interaction, to coupled-resonator optical waveguides (CROWs), in which slow-light can be used to enhance the nonlinear interaction \cite{Melloni2003365,davanco2012}. Yet, the tuning and control of a system composed of many linearly coupled resonators can be challenging, with the normal modes of the system typically extended over the full size of the structure, or at least over many resonators. Acting on a single element of the structure can have erratic effects on the spectral properties of the whole system.

\begin{figure}[!b]
\centering
\includegraphics[width=\linewidth]{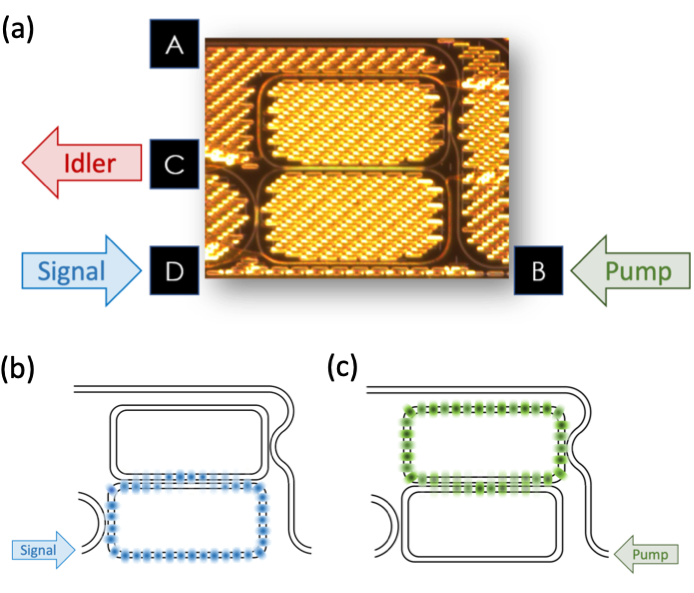}
\caption{(a) Optical micrograph of the sample. The arrows indicate the configuration used in the FWM experiment. (b) Schematic showing the typical field distribution when the structure is resonantly  excited from port D, as in the case of the Signal field. (c) Schematic showing the typical field distribution when the structure is resonantly excited from port B, as in the case of the Pump field.}
\label{fig:sample}
\end{figure}

In a recent work Menotti \emph{et al.} suggested a different approach, based on employing nonlinear coupling between linearly uncoupled normal modes of a structure \cite{Menotti2019}.
In the suggested implementation the linear modes of interest are mainly localized in different resonators, and the isolation is realized by a properly designed directional coupler (DC). Although the modes are linearly uncoupled, they share the DC region, where the nonlinear interaction can take place.
At first glance, this arrangement may seem disadvantageous, since it limits the spatial overlap region shared by the modes of the two parent resonators, thereby reducing the effective strength of the nonlinear processes. However, since the modes associated with the different resonators are linearly uncoupled, they can be tuned independently. This allows for the enhancement of quantum correlations of light generated in a dual-pump Spontaneous FWM configuration \cite{Menotti2019}, and the compensation of SPM and XPM. In this manner, the detrimental effects of nonlinear processes can be suppressed or compensated by a factor which can dramatically exceed the corresponding penalty on the efficiency of the desired nonlinear process.

In this work we show that this theoretical proposal can be implemented in practice, and we experimentally demonstrate four-wave mixing (FWM) in a system of two linearly uncoupled micro-resonators in which the pump field is resonantly coupled to the first resonator, while idler and signal fields are resonant with the second. The linear independence of the modes associated with the different resonators allows for on-chip optimization of the nonlinear process by tuning the resonances of one of the resonators so that pump, signal, and idler are all on resonance.

An optical micrograph of the structure is shown in Fig.~\ref{fig:sample}. The structure is composed of two Si$_3$N$_4$ racetracks embedded in silica that are adjacent and separated by a gap {$d=500$~nm}, forming a DC of length {$L=235.62$ $\mu$m}, with $d$ and $L$ chosen such that the two resonators are essentially linearly uncoupled in the frequency range of interest. In Fig. \ref{fig:sample}~(b) and (c) we show a sketch indicating the expected field distributions when the structure is pumped from port D and B, respectively. The modes of each racetrack are well-localized in space, with the DC being the only region shared by the modes belonging to different racetracks. The waveguide has width $w=1\ \mu$m and height $h=0.8\ \mu$m, while the top and bottom racetracks have total lengths {$\mathcal{L}_1=1.003$~mm} and  {$\mathcal{L}_2=1.023$~mm}, respectively. Each of the two resonators can be tuned independently by means of an electric heater located far from the DC to limit thermal cross-talk. The sample was fabricated by Ligentec SA \cite{ligentec}. 
\begin{figure}[!t]
\centering
\includegraphics[width=\linewidth]{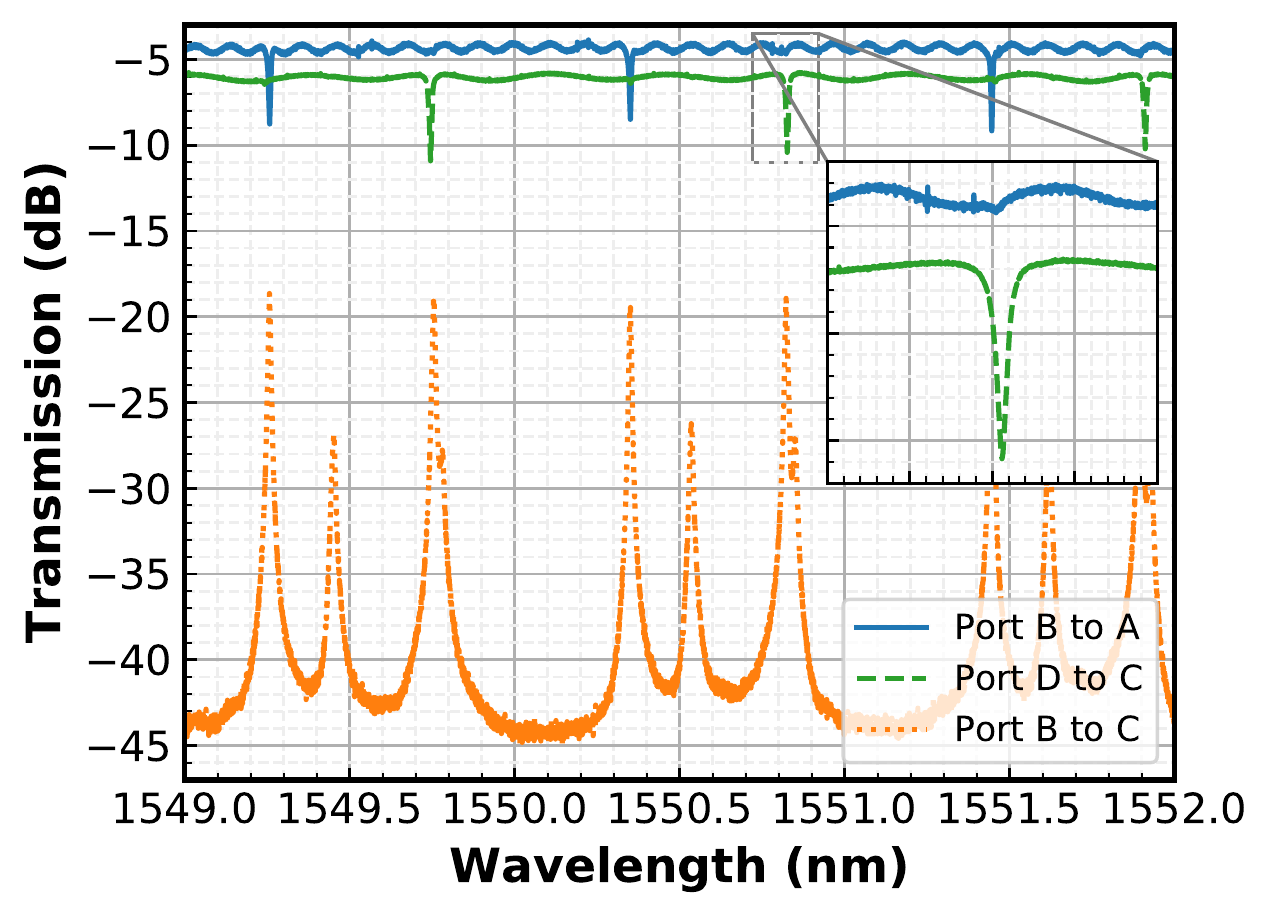}
\caption{Linear transmission spectra of the sample for various configurations: input and output from the same racetrack (port B to A, and port D to C), and input and output from different racetracks (port B to C). The input light excites mostly the fundamental TE mode of the waveguide, with a residual TM mode component. The inset on the top right shows a zoomed in racetrack resonance.}
\label{fig:transmission}
\end{figure}

We first characterized the linear properties of the structure to quantify the linear coupling between the two resonators. The fabricated chip was tested on chip coupling nano-stages, while lensed fibers were used for edge coupling of the light between standard SMF-28 optical fibers and the chip. We utilised a high resolution test system (Upgraded BOSA 400 with component analyzer) to sweep the laser wavelength, and measured the transmitted power in each port. The linear transmission spectra for various input/output configurations are shown in Fig. \ref{fig:transmission}. We measured high transmission for the B-A and D-C combinations, which respectively correspond to input and output ports both belonging to the upper or lower resonators. Each spectrum shows the resonances of the fundamental quasi-transverse-electric (TE) modes of the corresponding racetracks. For both resonators, we measure a loaded quality factor (Q) of about 200,000. Based on the line width and extinction of the resonances extracted by fitting those spectra with a Lorentzian function, the resonators are estimated to be overcoupled with an escape efficiency of $\varepsilon=79\%$ and an intrinsic Q of about 1 million. When considering input and output ports belonging to different resonators (the B-C case), the transmission is suppressed, with an isolation of about 10 dB on-resonance and up to 40 dB off-resonance. Such isolation is affected by fabrication variations of the material index and the DC geometry. For instance, other devices on the same wafer presented better linear isolation; however, the position of the resonances in those devices was beyond the heater tuning range to achieve FWM. A number of solutions could be envisioned to improve the isolation: one can for instance make use of tunable DCs, in which fabrication variations could be actively compensated by thermal tuning \cite{Orlandi13}.

In the B-C transmission spectrum one can observe all the resonances of the two resonators, including those corresponding to fundamental quasi-transverse-magnetic (TM) modes, which are much weaker and visible in the B-C case because of a residual TM mode coming from imperfect fiber-to-chip polarization alignment. We note that the transmission spectra have not been corrected for the fiber-to-chip coupling losses and chip waveguide losses, which give total insertion losses of 4 dB and 6 dB for port B to A and port D to C, respectively.

\begin{figure}[!t]
\centering
\includegraphics[width=\linewidth]{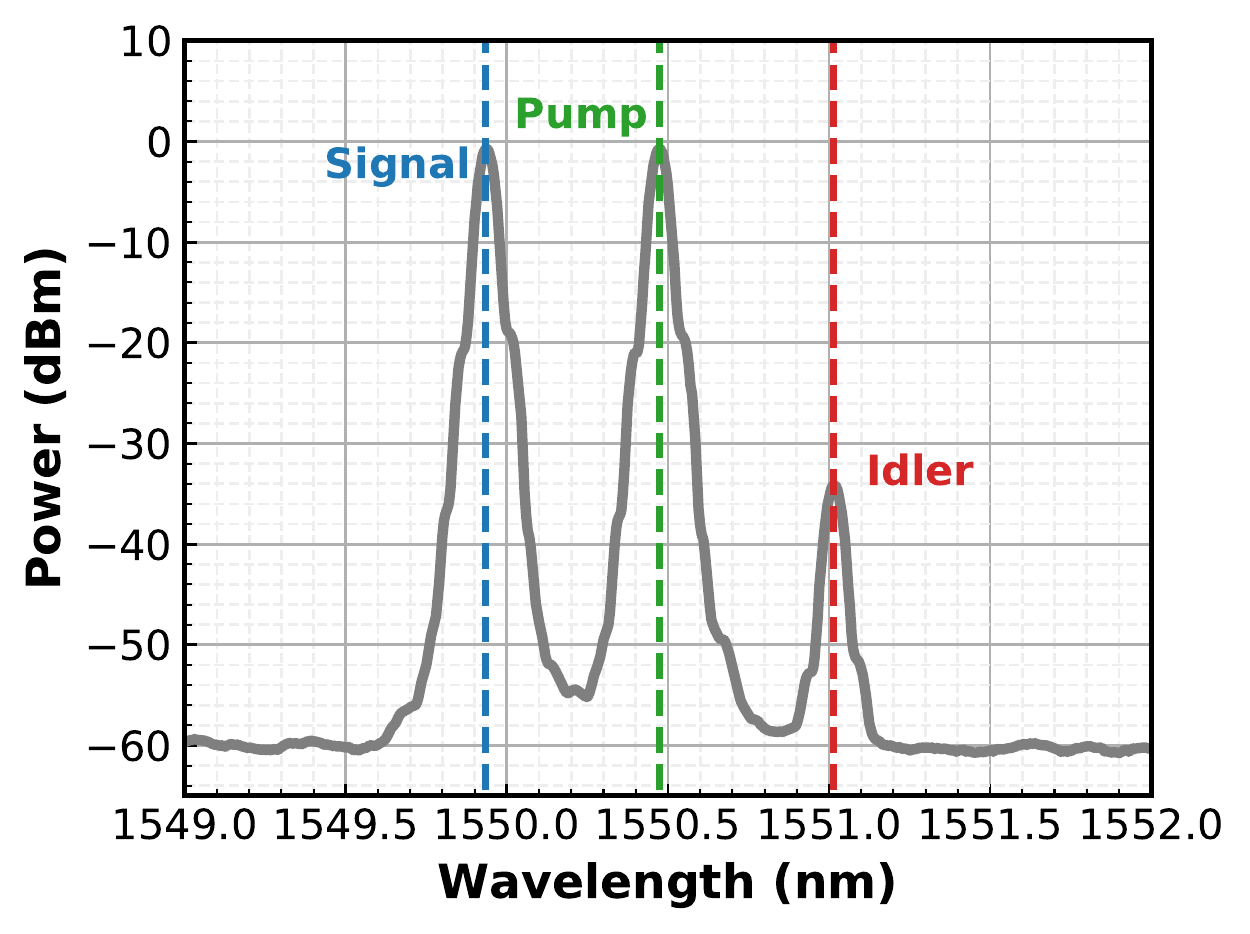}
\caption{Unfiltered optical power measured at port C in the case of a triply resonant configuration; the spectral position of signal, pump, and idler resonances are indicated in blue, green and red dashed lines, which are tuned to fulfill the energy conservation condition for FWM.}
\label{fig:nonlinear}
\end{figure}


Although the two resonators are nearly linearly uncoupled, one can observe FWM generated in the DC region shared by the modes of both resonators. As usual, efficient FWM requires that the photon energies of signal, pump, and idler must be equally spaced. In a single resonator, when group velocity dispersion (GVD) is negligible, the resonances are equally spaced only in the limit of low pump intensity. Indeed, at high pump intensities, SPM and XPM induce uneven shifts of the resonances, and equal spacing is lost. To compensate for these effects, one must start with a dispersion-engineered structure for which the resonance separations in energy are unequal at low intensities \cite{Pfeiffer2017684}. Since this involves engineering the spectral positions of the resonances with a precision comparable with their linewidth, having three equally spaced resonances in a single resonator at the operating point can be challenging for high-Q resonators. However, in our structure we can operate by using a pump on resonance in one of the resonators, and the signal and idler on resonance in the other. Since the resonances of each resonator can be tuned independently by local thermal heating via integrated micro-heaters, the comb of one of the resonators can be adjusted to give equal spacing between the three resonances of interest, regardless of the GVD  or additional power-dependent energy shifts due to SPM or XPM. Here we consider the configuration sketched in Fig. \ref{fig:sample}, in which the pump is injected in the top resonator through port B and the signal is input in the bottom resonator through port D; the generated idler light is collected from port C.

\begin{figure}[!t]
\centering
\includegraphics[width=\linewidth]{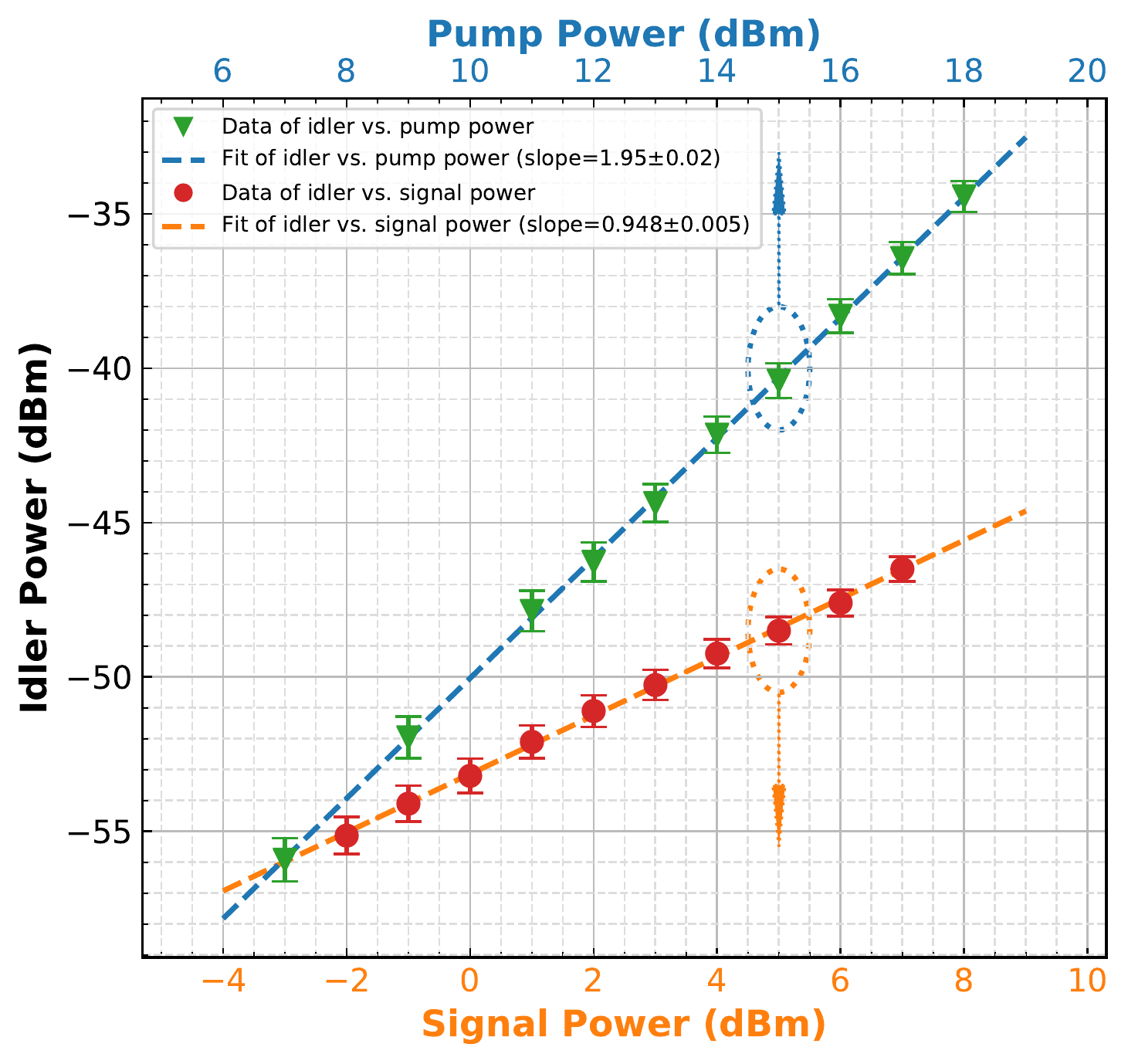}
\caption{Measured idler power as a function of the pump (triangles) and signal (circles) input power. Dashed lines correspond to linear fits of the data, demonstrating quadratic and linear scaling, respectively.}
\label{fig:scaling}
\end{figure}

In Fig.~\ref{fig:nonlinear} we show the spectrum measured from port C using an Agilent 86142B optical spectrum analyzer (OSA), where we also indicate the position of the three resonances that have been tuned to be equally spaced in energy. When we inject a CW pump tuned at $1550.475$~nm with an on-chip power $P_P=63$~mW into port B and a CW signal tuned at $1549.935$~nm with an on-chip power $P_S=5$~mW into port D, we measure a generated idler at $1551.015$~nm with an off-chip power of $0.3864\ \mu$W. Although the spectrum has been measured without any spectral filter, the pump is suppressed because it is injected in the top resonator while the idler is generated on resonance with the bottom resonator and collected from port C. The pump filtering effect is a direct consequence of having the two resonators linearly uncoupled. Finally, it should be noted that we are working with CW inputs, with the output spectrum reported in Fig.~\ref{fig:nonlinear} appearing broader than the laser linewidth only because of the resolution of the OSA used in the nonlinear measurements.

Based on the data in Fig.~\ref{fig:nonlinear}, we can calculate the on-chip FWM efficiency $\eta_{exp}=P_I/({\alpha P_S})\simeq 1.54\times 10^{-4}$ at $P_P=63$~mW, where $\alpha\simeq 0.5$ ($-3$~dB) is the coupling efficiency to the chip. 
In the ideal case of perfect pump suppression from the DC and no signal or pump detuning, and with modest overcoupling, the theoretical value of the on-chip FWM conversion efficiency is \cite{,Menotti2019,Absil:00} 
\begin{equation}\label{FWM_theory}
    \eta_{th}=\Big(\gamma P_P \frac{L}{4}\Big)^2\Big(\frac{2}{\pi}\Big)^4 \mathcal{F}_P^2 \mathcal{F}_S \mathcal{F}_I,
\end{equation}
where $\gamma$ is the waveguide nonlinear parameter, which we estimate in $\simeq 1\ \mathrm{m^{-1}W^{-1}}$, $\mathcal{F}_{P}$ is the finesse of the top resonator at the pump frequency, and  $\mathcal{F}_{S(I)}$ is the finesse of the bottom resonator at the signal (idler) frequency.
Based on \eqref{FWM_theory} and ignoring the imperfections in our device design, fabrication, and experimental setup, the estimated conversion efficiency would be $\simeq 6.1\ 10^{-4}$, which is on the same order as the experimental data.

Once momentum and energy conservation are satisfied, FWM can occur wherever the spatial overlap of the modes involved is not null. This condition is certainly verified on the DC, but a contribution to the measured idler intensity can also come from light generated by FWM in the bottom resonator due to the residual pump transmission. This process is expected to be weak, for the bottom resonator is not resonant at the pump frequency, and thus the pump field in the bottom ring is strongly attenuated. The ratio between the idler power $P_{I,\mathrm{DC}}$ generated in the directional coupler and the idler power $P_{I,\mathrm{Ring}_\delta}$ generated in the bottom ring when the pump is of resonance by $\delta$ is  
\begin{equation}\label{SNR}
\frac{P_{I,\mathrm{DC}}}{P_{I,\mathrm{Ring}_\delta}}=\frac{1}{T^2}\left(\frac{L}{4\mathcal{L}_2}\right)^2\left(\frac{\delta^2+\Delta^2}{\Delta^2}\right)^2,    
\end{equation}
where $T$ is the pump transmission measured in port C, $\Delta$ is the FWHM of resonances of the bottom ring, and $\delta$ is the detuning from the resonant condition. In our case, considering that $\delta\cong FSR/2$ we estimate that the ratio in \eqref{SNR} is about 1600. Therefore, we can safely ignore the nonlinear interaction taking place outside the DC.

In order to demonstrate that the generated light originates from a nonlinear interaction, specifically FWM, we show in Fig.~\ref{fig:scaling} the measured idler power as a function of pump and signal powers. The dashed lines are linear fits for the experimental data of measured idler power as a function of signal and pump power, whose best-fit slopes are $0.948\pm0.005$  and $1.95\pm0.02$, respectively. The two curves have the expected linear and quadratic scaling with $P_S$ and $P_P$, respectively, which is a clear signature of FWM. The scaling with the pump power is quadratic over the entire range considered here (up to 18~dBm), confirming that the nonlinear losses are negligible, as expected for Si$_3$N$_4$.

\begin{figure}[!t]
\centering
\includegraphics[width=\linewidth]{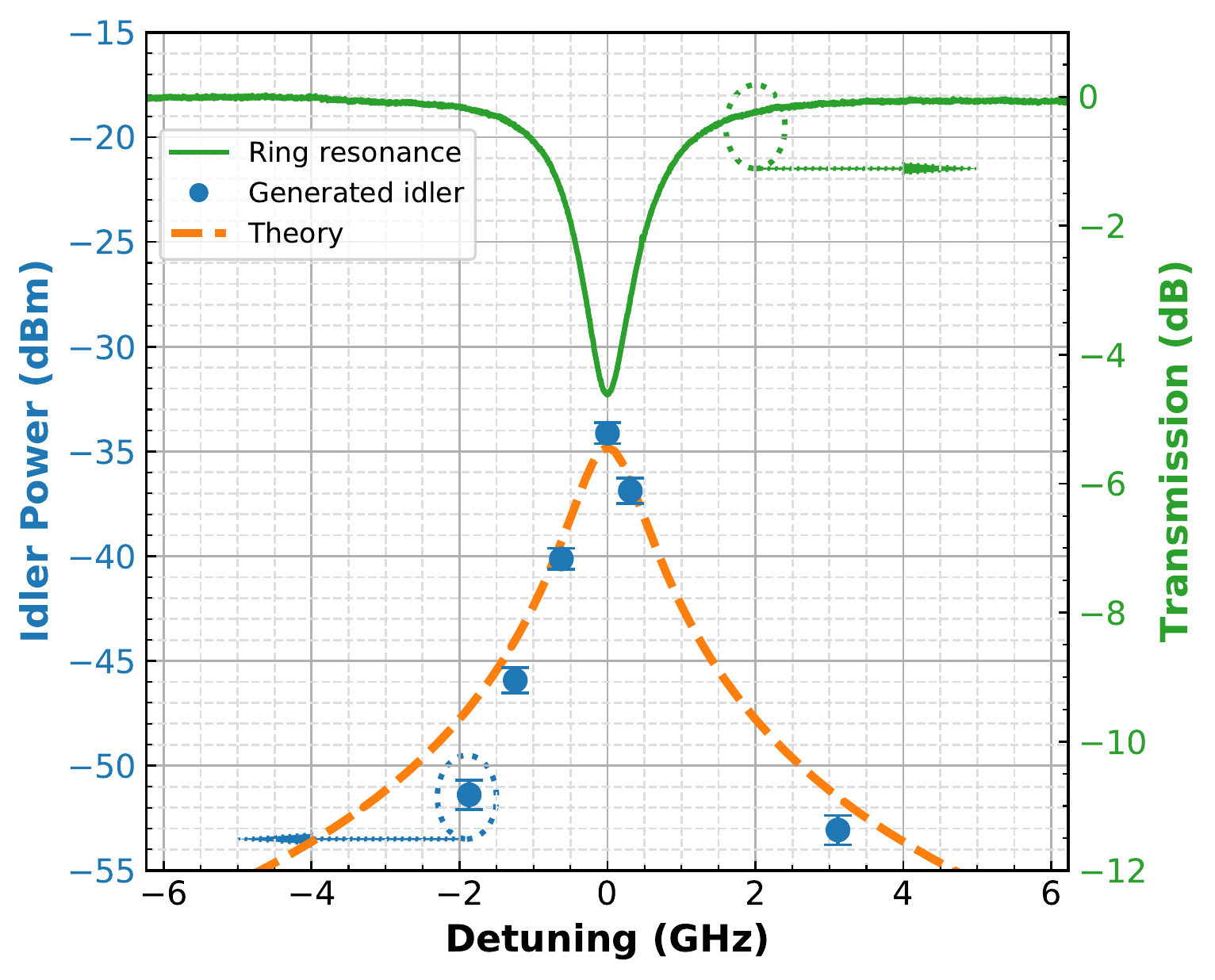}
\caption{Measured idler power as a function of the detuning of the idler resonance from the position corresponding to the triply resonant condition (circle). Linear transmission of the bottom racetrack in the case of triply resonant condition (solid green). Theoretical prediction of the idler attenuation due to the resonance detuning (orange dashed).}
\label{fig:detuning}
\end{figure}

Finally, we confirm the capability of controlling the strength of FWM occurring in the DC by adjusting the heaters to tune the relative position of the two resonance combs, one associated with each resonator.  Specifically, we resonantly couple a CW pump to the top resonator through port $B$, with the signal field injected in port $D$ and kept on resonance while the comb of the bottom racetrack is gradually shifted. The measured idler power is shown in Fig.~\ref{fig:detuning} as a function of the detuning from the configuration corresponding to the case of three equally-spaced resonances. In this configuration presented  in Fig.~\ref{fig:nonlinear}, the condition of energy conservation $2\omega_P-\omega_S-\omega_I$ is satisfied with all the fields on resonance, and the nonlinear coupling is the strongest. When the comb of the bottom racetrack is detuned by $\delta$ (either positively or negatively from that of the top racetrack) and the signal field is kept on resonance, the generated idler field is no longer on resonance. Specifically, the idler is generated off-resonance by $2\delta$, thus the conversion efficiency is reduced by a factor $\Delta^2/(\Delta^2+4\delta^2)$, as shown by theoretical curve Fig.~\ref{fig:detuning}, which is in good agreement with the experimental data.

In conclusion, we experimentally demonstrated the nonlinear coupling of linearly uncoupled resonators via stimulated FWM in a system of two coupled racetracks. In this structure the comb of resonances associated with one racetrack can be tuned independently of the comb of resonances associated with the other racetrack. This independent tuning allows for the condition of equally spaced resonances to be achieved, even in the presence of strong SPM and XPM, by on-chip adjustment of the racetrack heaters. Thus, efficient FWM can be realized without the need for dispersion engineering. Moreover, linearly uncoupled resonators can provide an additional layer of pump suppression, which is typically desired in nonlinear processes.
This approach is robust and flexible, and it can be applied to other nonlinear interactions such as spontaneous FWM or coherent Raman scattering. Therefore, we believe it will be particularly useful and practical for the on-chip control of other nonlinear optical phenomena.

\setlength{\parskip}{1em}
The authors thank L. G. Helt for his critical reading of the manuscript and useful comments.

\bibliography{biblio}

\end{document}